\documentclass[10pt,journal,final,twocolumn]{IEEEtran}

\usepackage{graphics}
\usepackage{graphicx}
\usepackage{subfigure}
\usepackage{epstopdf}
\usepackage{epsfig}
\usepackage[cmex10]{amsmath}
\usepackage{amssymb}
\usepackage{times}
\usepackage{ntheorem}
\usepackage{pifont}
\usepackage{stfloats}
\usepackage{bm}
\usepackage{color}
\usepackage{makecell}
\usepackage{array}
\usepackage{multirow}
\usepackage{booktabs}

\begin{document}
%
\title{\huge Pruned Convolutional Attention Network Based Wideband Spectrum Sensing with Sub-Nyquist Sampling}

\author{\IEEEauthorblockN{
Peihao~Dong,~\IEEEmembership{Member,~IEEE}, Jibin~Jia, Shen~Gao, Fuhui~Zhou,~\IEEEmembership{Senior Member,~IEEE}, \\
and Qihui Wu,~\IEEEmembership{Fellow,~IEEE}
}

\vspace{-0.4cm}
\thanks{
\vspace{-0.3cm}


This work was supported in part by the National Science Foundation of China under Grants 62101253 and 62471226, the Natural Science Foundation of Jiangsu Province under Grant BK20210283, the open research fund of National Mobile Communications Research Laboratory, Southeast University under Grant 2022D08, and the Fundamental Research Funds for the Central Universities under Grant NS2024023. \emph{(Corresponding author: Peihao Dong.)}

P. Dong is with College of Electronic and Information Engineering, Nanjing University of Aeronautics and Astronautics, Nanjing 211106, China, and also with the National Mobile Communications Research Laboratory, Southeast University, Nanjing 211111, China (e-mail: phdong@nuaa.edu.cn).

J. Jia, F. Zhou, and Q. Wu are with College of Electronic and Information Engineering, Nanjing University of Aeronautics and Astronautics, Nanjing 211106, China (e-mail: jiajibin@nuaa.edu.cn; zhoufuhui@ieee.org; wuqihui2014@sina.com).

S. Gao is with the College of Telecommunications and Information Engineering, Nanjing University of Posts and Telecommunications, Nanjing 210003, China (e-mail: gaoshen\_ee@163.com).

}
}

\IEEEtitleabstractindextext{%
\begin{abstract}
Wideband spectrum sensing (WSS) is critical for orchestrating multitudinous wireless transmissions via spectrum sharing, but may incur excessive costs of hardware, power and computation due to the high sampling rate. In this article, a deep learning based WSS framework embedding the multicoset preprocessing is proposed to enable the low-cost sub-Nyquist sampling. A pruned convolutional attention WSS network (PCA-WSSNet) is designed to organically integrate the multicoset preprocessing and the convolutional attention mechanism as well as to reduce the model complexity remarkably via the selective weight pruning without the performance loss. Furthermore, a transfer learning (TL) strategy benefiting from the model pruning is developed to improve the robustness of PCA-WSSNet with few adaptation samples of new scenarios. Simulation results show the performance superiority of PCA-WSSNet over the state of the art. Compared with direct TL, the pruned TL strategy can simultaneously improve the prediction accuracy in unseen scenarios, reduce the model size, and accelerate the model inference.
\end{abstract}

}

\maketitle

\IEEEdisplaynontitleabstractindextext

%
\IEEEpeerreviewmaketitle

\vspace{-0.2cm}
\section{Introduction}

\IEEEPARstart{W}{ith} the emergence of new devices and applications, the wireless network is anticipated to evolve into a versatile architecture spanning over a ultra-broad frequency range \cite{Q. Wu}--\cite{P. Dong}. It is of great importance to orchestrate the multitudinous wireless transmissions in a dynamic access manner and thus the effective wideband spectrum sensing (WSS) is necessary \cite{J. Fang}.

In distinction to the narrowband case, Nyquist sampling rate proportional to the bandwidth becomes impractical for the WSS since it requires the more complex hardware and higher power consumption. Besides, the mass of sampled data will aggravate the computation load of signal processing. Then lots of researches focus on the reliable WSS under sub-Nyquist rates. As one of the representative solutions therein, compressed sensing (CS) based approaches treat the spectrum occupancy detection as a reconstruction problem of the sparse signal from limited sampled data \cite{S. K. Sharma}. In \cite{Z. Qin}, both the single and cooperative WSS approaches were developed based on CS and then validated using real-world data. An iterative compressive filtering algorithm was proposed in \cite{W. Xu} to successively detect the occupied frequency band without the signal recovery.

Despite some highlights, CS-based approaches may suffer from performance deterioration in practical scenarios with the non-ideal sparse property, e.g., high occupancy ratio of the spectrum. The revival of deep learning (DL) provides potential solutions for these problems due to its unique ability of handling complicated, non-linear tasks, as shown by its successful applications in wireless signal classification \cite{T. J. O'Shea}, resource allocation \cite{H. Ye}, transceiver design \cite{H. He}, and so on. In \cite{J. Zhang}, the convolutional neural network (CNN) is applied to conduct the multiband occupancy detection based on the covariance matrices. A WSS framework consisting of two parallel, small-size CNNs was proposed in \cite{R. Mei}, where the outputs of two branches are fused to improve the detection accuracy. To accommodate the sub-Nyquist rate, a fully-connected (FC) neural network based spectrum sensing approach was designed by using the multicoset sampled data in \cite{H. Zhang}. In \cite{S. Chandhok}, a series of neural networks are developed for WSS and modulation classification under the sub-Nyquist rate. In \cite{X. Zhang}, DL is exploited to modify the alternating direction method of multipliers network for the wideband signal reconstruction.

While inheriting the advantages of DL, the DL-based WSS approaches are challenged by the generalization and complexity problems plaguing most DL applications. To tackle these problems, we develop an attention mechanism based WSS framework with the improved robustness in a low-cost manner. The main novelty and contribution can be summarized as follows:

\begin{itemize}[\IEEEsetlabelwidth{Z}]
\item[1)] The multicoset preprocessing and convolutional attention mechanism are orchestrated to enable the sub-Nyquist sampling while still achieving the superior sensing accuracy. Specifically, the embedding procedure of the attention mechanism is absorbed into the multicoset preprocessing, which provides a tentatively estimated spectrum of the wideband signal as an initial feature representation. Different from the regular multiplication way to generate the attention query, key, and value matrices, the convolution operation is adopted in order to process the three-dimensional (3D) feature matrix. Weight pruning is applied to reduce the model complexity, yielding a low-cost pruned convolutional attention WSS network (PCA-WSSNet) almost without the performance loss.

\item[2)] A transfer learning (TL) strategy for PCA-WSSNet is developed to cope with the highly varied spectrum environment by adapting the domain-specific layers with a few adaptation samples. The model pruning further benefits the TL procedure since it reduces the number of weights need to be adapted considerably, leading to the better TL performance yet the lower complexity.

\item[3)] The simulation results show that the proposed PCA-WSSNet can maintain the high prediction accuracy regardless of the occupancy ratio. It also remarkably reduces the DL model complexity with the superior generalization capability. Therefore, PCA-WSSNet is a sound solution to the WSS problems due to the elaborate design of the network architecture and model pruning strategy.
\end{itemize}


\vspace{-0.3cm}
\section{System Model}

\begin{figure}[t]
	\centering
	\includegraphics[width=2.5in]{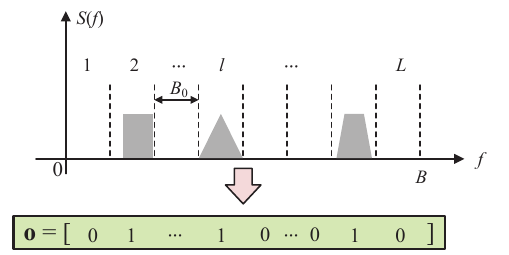}
	\caption{Occupancy model of the wideband spectrum.}\label{occupation model}
	\vspace{-0.3cm}
\end{figure}

Consider a WSS system consisted of $K$ primary users (PUs) and a cognitive radio (CR) receiver for purpose of opportunistic spectrum access or spectrum surveillance. As shown in Fig.~\ref{occupation model}, the wideband spectrum of interest ranges from 0 to $B$ and is evenly divided into $L$ non-overlapped sub-bands with the bandwidth $B_{0}=B/L$ and the index set $\mathcal{I}=\{1,\dots,L\}$. Considering that the signal of each PU occupies one sub-band, the spectrum occupancy situation can be indicated by an $L$-entry binary vector, $\mathbf{o}$, composed of $K$ non-zero elements, which needs to be estimated by the CR receiver based on the received signal.

During the spectrum sensing period, the received signal compounding $K$ concurrent PU signals at the CR receiver is given by
\vspace{-0.3cm}
\begin{eqnarray}
	\label{eqn_xt}
	x(t) = \sum_{k=1}^{K} s_{k}(t) + z(t),
\end{eqnarray}
where $s_{k}(t)$ and $z(t)\sim\mathcal{CN}(0,\sigma^2)$ denote the signal from the $k$th PU and the additive white Gaussian noise (AWGN), respectively. Using Fourier transform on $x(t)$ yields $X(f) = \int_{-\infty}^{\infty} x(t) e^{-j2\pi f t} dt  = S(f) + Z(f)$,
where $S(f)$ and $Z(f)$ denote the spectrums of the compound PU signal and AWGN, respectively.

Since the spectrum occupancy situation varies over the time, the periodical spectrum sensing is needed. The traditional Nyquist sampling rate proportional to the bandwidth will incur the high hardware cost and power consumption, and is not suitable for the WSS that is an important task in the future wireless networks. Then we aim to design an efficient WSS scheme under the sub-Nyquist rate by resorting to DL in the following sections.

\section{PCA-WSSNet Design}

In this section, the proposed PCA-WSSNet is designed. The multicoset preprocessing (MP) with sub-Nyquist sampling is first elaborated, followed by detailing the backbone architecture and pruning procedure of PCA-WSSNet.

\subsection{Multicoset Preprocessing}

Multicoset sampling adopts several parallel branches, called coset, to respectively digitize the received analog signal at a rate lower than the Nyquist rate $f_{\textrm{N}}$. For the $p$th branch of a $P$-coset sampling system, using the sampling rate $f_{\textrm{b}}=f_{\textrm{N}}/L=1/LT$ yields the discrete-time sequence expressed as \cite{M. Mishali}
\vspace{-0.1cm}
\begin{eqnarray}
	\label{eqn_y_p}
	y_p[n] = x(nLT+c_p T), \quad n\in \mathbb{Z},
\end{eqnarray}
where $T$ denotes the Nyquist sampling interval and $c_p$ is an integer controlling the time shift of the $p$th branch. With the set $\mathcal{C}=\{c_1,\ldots,c_P\}$ satisfying $0 \leq c_1<\cdots< c_P \leq L-1$, the average sampling rate of the $P$-coset sampling system is $f_{\textrm{ave}}=\frac{P}{LT}=\frac{P}{L}f_{\textrm{N}}$, revealing that $f_{\textrm{ave}}<f_{\textrm{N}}$ holds true so long as $P<L$ is set.

Calculating the discrete-time Fourier transform of $y_p[n]$ along with some manipulations derives $Y_p(e^{j2\pi fLT})=e^{j2\pi f c_p T} \sum_{l=1}^{L}\frac{e^{-j2\pi (l-1) c_p/L}}{LT}X(f-\frac{l-1}{LT})$. Then a compact matrix form can be given as $\mathbf{y}(f)=\mathbf{A} \mathbf{x}(f)$ with their entries $y_p(f)=e^{-j2\pi f c_p T} Y_p(e^{j2\pi fLT})$, $x_{l}(f)=X(f-\frac{l-1}{LT})$, and $A_{p,l}=\frac{e^{-j2\pi (l-1) c_p/L}}{LT}$, respectively, for $f \in [0,B/L)$, $p=1,\ldots,P$ and $l=1,\ldots,L$. Then a coarse estimate of $\mathbf{x}(f)$ is given by

\vspace{-0.2cm}
\begin{eqnarray}
	\label{eqn_x_vec_hat}
	\hat{\mathbf{x}}(f)=\mathbf{A}^{\dagger}\mathbf{y}(f)=\mathbf{A}^{\dagger}\mathbf{A} \mathbf{x}(f),
\end{eqnarray}
where $(\cdot)^{\dagger}$ denotes the pseudo-inverse. Considering that each of $P$ sampled sequences consists of $N$ data, values of $\hat{\mathbf{x}}(f)$ at $N$ frequencies can be obtained to construct a matrix $\hat{\mathbf{X}}(f)=[\hat{\mathbf{x}}(f_0),\ldots,\hat{\mathbf{x}}(f_{N-1})]$, which is then normalized as $\bar{\mathbf{X}}=\frac{\hat{\mathbf{X}}(f)}{|\hat{\mathbf{X}}(f)|}$. Note that the processing in (\ref{eqn_x_vec_hat}) can be regarded as the embedding operation of the attention mechanism. In more detail, (3) acts two roles: Estimating the spectrum of the received signal in the MP and generating the initial feature matrix that is tractable for the subsequent convolutional attention (CA) network, making it a shared step smoothly connecting the MP and CA network.


\vspace{-0.3cm}
\subsection{CA-WSSNet Architecture}

\begin{figure*}[t]
	\centering
	\includegraphics[width=4.8in]{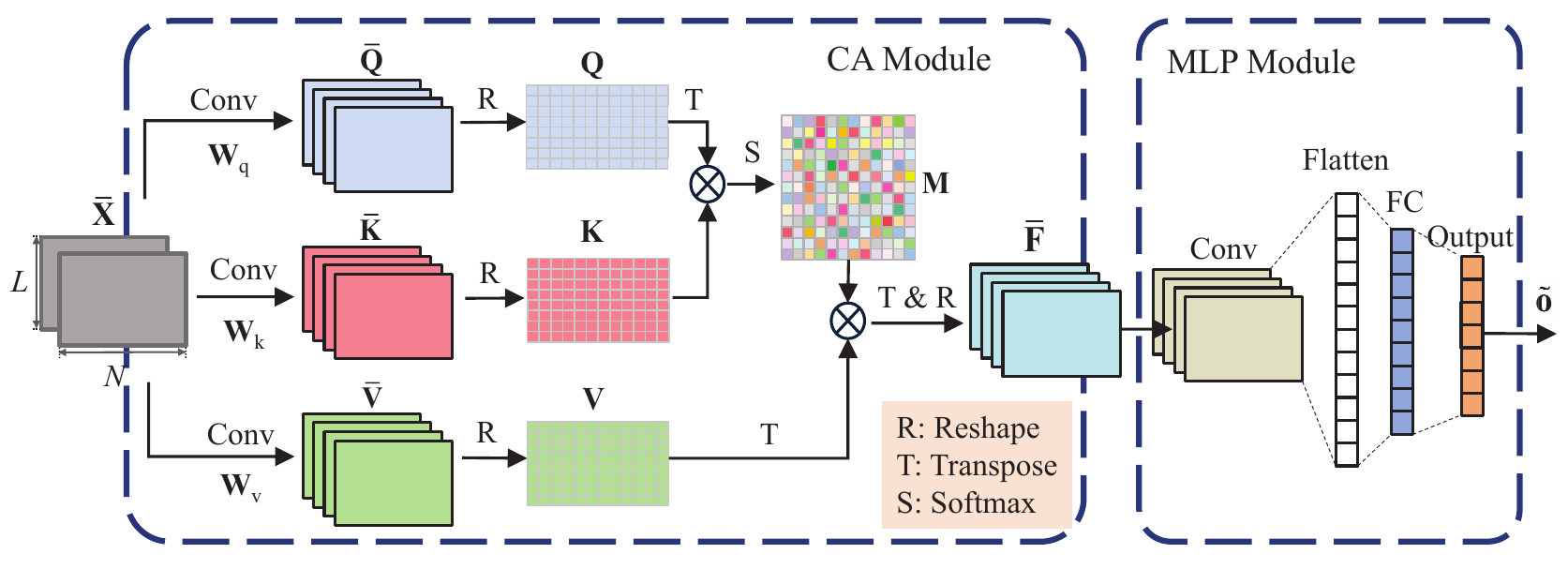}
	\caption{CA-WSSNet Architecture.}\label{WSSNet}
	\vspace{-0.3cm}
\end{figure*}

In this subsection, the architecture of CA-WSSNet is designed. The FC, convolutional, and recurrent structures have been widely adopted to design the spectrum sensing network, which, however, neglect the importance difference of the elements in the feature matrix. To effectively capture the importance difference for the more accurate prediction, a CA module is designed first. Then a multi-layer perceptron (MLP) module is appended to output the sensing result. It is noted that the convolutional (Conv) layer extracts features only within a local region while the attention mechanism processes the feature maps more comprehensively to allocate the information importance scores for ease of the subsequent processing. So it is preferable to place the CA module in front, followed by the Conv layer of the MLP module to further capture the key features correlated to the output. As per simulation trails, this placement sequence of modules achieve the best performance. The detailed architecture of CA-WSSNet is illustrated in Fig.~\ref{WSSNet}, which receives the initial feature matrix $\bar{\mathbf{X}}$ as the input to approximate the true occupancy vector $\mathbf{o}$. In the CA module, different from the regular multiplication way to generate the query, key, and value matrices, the convolution operation is applied so that the 3D feature matrix $\bar{\mathbf{X}}\in \mathbb{R}^{L \times N \times 2}$ can be effectively processed as
\vspace{-0.1cm}
\begin{eqnarray}
	\begin{aligned}
		\label{eqn_qkv}
		\bar{\mathbf{Q}}=\mathbf{W}_{\mathrm{q}}\ast\bar{\mathbf{X}}, \bar{\mathbf{K}}=\mathbf{W}_{\mathrm{k}}\ast\bar{\mathbf{X}}, \bar{\mathbf{V}}=\mathbf{W}_{\mathrm{v}}\ast\bar{\mathbf{X}},
	\end{aligned}
\end{eqnarray}
where $\mathbf{W}_{\mathrm{q}}$, $\mathbf{W}_{\mathrm{k}}$, and $\mathbf{W}_{\mathrm{v}}$ denote the zero-padding (ZP) convolutional kernels with size $3\times 3$ and four output channels used to respectively generate $\bar{\mathbf{Q}}, \bar{\mathbf{K}}, \bar{\mathbf{V}}\in \mathbb{R}^{L \times N \times 4}$. Then $\bar{\mathbf{Q}}$, $\bar{\mathbf{K}}$, and $\bar{\mathbf{V}}$ are reshaped by horizontally stacking four feature maps so that the query, key, and value matrices $\mathbf{Q}, \mathbf{K}, \mathbf{V}\in \mathbb{R}^{L \times 4N}$ can be obtained. Multiply $\mathbf{Q}^T$ and $\mathbf{K}$ in the scaled dot-product manner and then apply the Softmax function to generate the importance matrix, that is
\vspace{-0.2cm}
\begin{eqnarray}
	\label{eqn_sftmx}
	\mathbf{M} = \mathrm{Softmax}\left(\frac{\mathbf{Q}^T\mathbf{K}}{\sqrt{L}}\right).
\end{eqnarray}
The importance matrix $\mathbf{M}\in \mathbb{R}^{4N \times 4N}$, also called the attention map, with each element in the interval $[0,1]$ actually reveals the importance scores of $4N$ input features versus $4N$ output features and acts on the value matrix as
\vspace{-0.1cm}
\begin{eqnarray}
	\label{eqn_MV}
	\mathbf{F}=\mathbf{M}\mathbf{V}^T.
\end{eqnarray}
Then $\mathbf{F}^T\in \mathbb{R}^{L \times 4N}$ is inversely reshaped by splitting the second dimension to output the feature map $\bar{\mathbf{F}}\in \mathbb{R}^{L \times N\times 4}$, which is then processed by the MLP module consisted of one Conv layer, one FC layer, and the output layer of CA-WSSNet. The Conv layer uses four kernels with size $3\times 3$ to enhance the feature representation. The FC layer synthesizes the flattened feature map into a task-specific feature vector with length $128$. Both the Conv layer and FC layer apply rectified linear unit (ReLU) activation function. Finally, the output layer transforms the feature vector into the probability vector $\tilde{\mathbf{o}}$ by using Sigmoid activation function. As the collocation of Sigmoid, the binary cross-entropy is adopted as the loss function to measure the discrepancy between the prediction $\tilde{\mathbf{o}}$ and the true label $\mathbf{o}$ in an element-wise manner, that is,
\vspace{-0.2cm}
\begin{eqnarray}
	\label{eqn_loss}
	J = -\frac{1}{N_{\textrm{tr}}}\sum_{n=1}^{N_{\textrm{tr}}}\sum_{l=1}^{L} o_{l}^{(n)}\log \tilde{o}_{l}^{(n)} + (1-o_{l}^{(n)})\log (1-\tilde{o}_{l}^{(n)}),
\end{eqnarray}
where $N_{\textrm{tr}}$ denotes the number of training samples and the superscript $(n)$ means the $n$th sample.

\vspace{-0.2cm}
\subsection{Weight Pruning of PCA-WSSNet}

To reduce the complexity of CA-WSSNet for the faster inference, the weights having trivial contributions to the prediction performance therein can be pruned, yielding PCA-WSSNet. The absolute value is a widely used metric to measure the contribution of the model weight, based on which the weights with absolute values below a predefined threshold are removed. To reduce the complexity while maintain the performance, CA-WSSNet will be partially pruned. Since the CA module, Conv layer, and FC layer are hidden layers and contain numerous weights, exerting weight pruning on them can significantly reduces the model complexity. In contrast, the output layer has only a few weights while acts the key role of transforming the feature vector into the final probability vector, meaning that the output layer is not appropriate to be pruned.

For ease of description, the CA module, Conv layer, FC layer, and output layer are indexed as Layer-$1,2,3,4$, respectively. Denote $\boldsymbol{\theta}_{i}$ as the vector composed of all $M_i$ weights of the $i$th layer for $i=1,2,3,4$. Then we focus on the pruning of $\{\bm{\theta}_{1}, \bm{\theta}_{2}, \bm{\theta}_{3}\}$. Given the pruning ratio $\kappa$, the pruning threshold $\gamma_i$ can be determined. For $i=1,2,3$, by sorting absolute values of the $M_i$ weights in an ascending order and putting them into a vector $\bm{\eta}_i$, $\gamma_i$ is given by $\gamma_i = [\bm{\eta}_i]_{\lceil\kappa{M_3}\rceil}$, where $\lceil\cdot\rceil $ represents the ceiling function and $[\bm{\eta}_i]_j$ denotes the $j$th element of $\bm{\eta}_i$. Then the weights in $\bm{\theta}_i$ are processed as
\begin{align}
	\label{eqn_prun}
	[\bm{\theta}_i]_j =
	\begin{cases}
		[\bm{\theta}_i]_j, &  |[\bm{\theta}_i]_j| \geq \gamma_i \\
		0, & \text{otherwise}
	\end{cases},
\end{align}
where $|\cdot|$ denotes the absolute value. Once a weight is set as $0$ as per (\ref{eqn_prun}), it becomes invalid and thus will not incur the computational cost in the subsequent model adaptation and inference stages. To compensate the performance loss caused by weight pruning, the remaining weights are fine-tuned by using the same datasets and loss function as the training stage, after which PCA-WSSNet is obtained.

\vspace{-0.2cm}
\section{Robust PCA-WSSNet Based on TL}

Due to the diversity of the PU type and the dynamics of the wireless environment, the spectrum occupancy situation is time-varying both in terms of the positions and the number of occupied sub-bands, posing the generalization problem for PCA-WSSNet. Under the fixed number of occupied sub-bands, the variation of their positions is relatively easy to generalize by varying the occupied positions in training data without need of covering all possible cases. However, this is problematic to handle the variation of the number of occupied sub-bands, or equivalently occupancy ratio, since the unseen cases will degrade the prediction accuracy. Including all cases into training data is also detrimental to the performance and is hard to realize as the total number of sub-bands becomes large. To tackle this problem, a TL strategy is developed to enhance the robustness of PCA-WSSNet when faced with new scenarios.

\begin{figure}[t]
	\centering
	\includegraphics[width=2.9in]{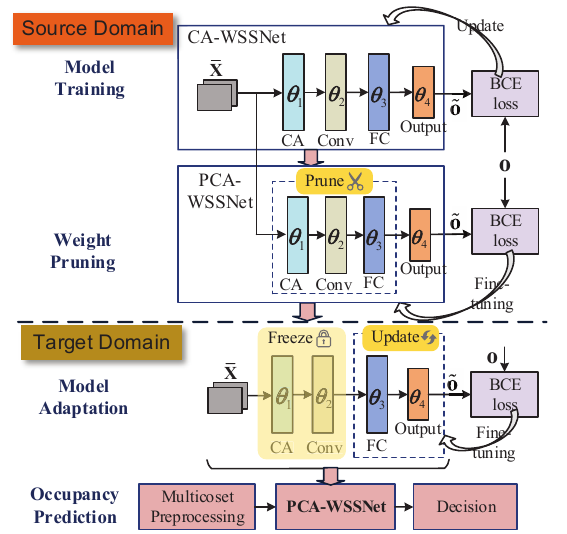}
	\caption{Workflow of PCA-WSSNet.}\label{PCA-WSSNet}
	\vspace{-0.3cm}
\end{figure}

\vspace{-0.2cm}
\subsection{Model Adaptation}

TL enables the DNN model to upgrade to a versatile predictor shuttling in the related tasks or domains via the appropriate weight adaptation, thereby detouring the difficulty of covering all possible cases in the offline-training stage \cite{J. Yosinsk}.

As illustrated in Fig.~\ref{PCA-WSSNet}, we aim to improve PCA-WSSNet trained in the source domain so that it is still effective in the unseen target domain. Generally, the neural layers of PCA-WSSNet can be divided into two parts, the part of general-feature layers and the part of domain-specific layers, which correspond to the layer-wise weight sets $\boldsymbol{\Theta}_{\mathrm{gf}}$ and $\boldsymbol{\Theta}_{\mathrm{ds}}$, respectively, with $\boldsymbol{\Theta}_{\mathrm{gf}}\cup \boldsymbol{\Theta}_{\mathrm{ds}}=\boldsymbol{\Theta}$. In the target domain, $\boldsymbol{\Theta}_{\mathrm{gf}}$ stays unchanged by transferring the general features from the source domain. On the other hand, due to the domain-dependent property, $\boldsymbol{\Theta}_{\mathrm{ds}}$ trained in the source domain is not suitable for the new target domain and thus needs the real-time update, that is,
\vspace{-0.2cm}
\begin{eqnarray}
	\label{eqn_update}
	\boldsymbol{\Theta}_{\mathrm{ds}}\gets \boldsymbol{\Theta}_{\mathrm{ds}}-\alpha\nabla J_{\mathcal{D}_{\mathrm{ad}}}(\boldsymbol{\Theta}_{\mathrm{ds}}),
\end{eqnarray}
where $\alpha$ denotes the learning rate for the gradient update and $\mathcal{D}_{\mathrm{ad}}$ denotes the dataset collected in the target domain for model adaptation. As shown in Fig.~\ref{PCA-WSSNet}, we have $\boldsymbol{\Theta}_{\mathrm{gf}}=\{\bm{\theta}_{1},\bm{\theta}_{2}\}$ and $\boldsymbol{\Theta}_{\mathrm{ds}}=\{\bm{\theta}_{3},\bm{\theta}_{4}\}$, i.e., fine-tune the FC and output layers. It is noted that the reasons of conducting the model pruning before the model adaptation are two-fold: 1) Model pruning can remarkably reduce the number of trainable weights so that the subsequent model adaptation can be accelerated. 2) Model pruning can be conducted only once offline, in order to avoid the frequent online execution consuming substantial resources.

Finally, PCA-WSSNet is used for the spectrum occupancy prediction by connecting the MP and decision procedures. Specifically, the decision procedure converts the output of PTL-WSSNet, $\tilde{\mathbf{o}}$, to the predicted binary occupancy vector, $\hat{\mathbf{o}}$, by comparing each element of $\tilde{\mathbf{o}}$ with a given threshold $\lambda \in (0,1)$ so that the spectrum occupancy situation can be revealed. The complete workflow of PCA-WSSNet is depicted in Fig.~\ref{PCA-WSSNet}.

\vspace{-0.2cm}
\subsection{Complexity Analysis}

In this subsection, the computational complexity of PTL-WSSNet in the model inference stage is analyzed in terms of the complexity of floating-point operations (FLOPs).

The computational complexity of CA-WSSNet is expressed as $\mathcal{O}\left(\sum_{i=1}^{2}C_{i}D_{i}M_{i}+\sum_{i=3}^{4}M_{i}+4N_{1}^2C_{1}D_{1}^2\right)$,
where $C_{i}$ and $D_{i}$ denote the length and width of convolutional feature maps of the $i$th layer, $N_1$ denotes the number of convolutional kernels of each branch in the CA module, where the subscript ``$1$" corresponds to the index of CA module, Layer-$1$. Note that the first two terms account for the weight-related computation while the remaining term corresponds to the matrix multiplications among $\mathbf{Q}, \mathbf{K}, \mathbf{V}$ in the CA module. Then the complexity of PCA-WSSNet can be written as $\mathcal{O}\left((1-\kappa)(\sum_{i=1}^{2}C_{i}D_{i}M_{i}+M_{3})+4N_{1}^2C_{1}D_{1}^2\right)$ by omitting the trivial item $M_{4}$. It can be seen that the weight pruning reduces a substantial part of the computational complexity.

\textbf{Remark 1 (Benefits from Weight Pruning):} \emph{Via the selective weight pruning, three main benefits can be reaped for PCA-WSSNet: 1) The model size can be reduced significantly along with the number of weights. 2) The model inference can be accelerated since the computational complexity is decreased. 3) The performance of model adaptation can be improved by reducing the number of trainable weights to better match the few adaptation samples, as revealed in following Fig.~\ref{Pacc_vs_Nad}.}

\vspace{-0.3cm}
\section{Simulation Results}

In this section, simulation results are provided to validate the effectiveness of the proposed PCA-WSSNet along with some insightful discussions. The compounded signal from $K$ PUs is generated as per
\begin{eqnarray}
	\label{eqn_st}
	s(t) = \sum_{k=1}^{K} \sqrt{E_kB_0} \operatorname{sinc}(B_0(t-t_k))e^{-j2\pi f_kt},
\end{eqnarray}
where $E_k=1$, $t_k\in{(0,T_d)}$ and $f_k\in{[B_0/2,B-B_0/2]}$ are the energy, time offset and carry frequency of $k$th PU's signal, respectively, with $T_d=8$ us and $\operatorname{sinc}(x)=\sin(\pi x)/(\pi x)$. Other key simulation parameters are set as follows. The total bandwidth is $B=320$ MHz and is evenly divided into $L=40$ sub-bands with the bandwidth $B_{0}=8$ MHz for each. For the multicoset sampling, the number of cosets is $P=16$ with $N=64$ sampled data for each branch. The threshold in the decision procedure is $\lambda=0.5$. For the DL models, the simulations are conducted on NVIDIA GeForce RTX 4060 graphical processing unit (GPU) with TensorFlow framework. The training, validation and testing sets contain $N_{\mathrm{tr}}=12000$, $N_{\mathrm{va}}=4000$, and $N_{\mathrm{te}}=4000$ samples. The pruning ratio is set $\kappa=0.9$ to achieve the good balance between the prediction accuracy and model complexity. For the offline training stage, the Adam optimizer is employed with the learning rate of $3\times 10^{-5}$ and the maximum number of epochs of $500$. The flag patience is set as $30$, meaning that the training will be terminated if the validation loss does not decrease for $30$ consecutive epochs. For the model pruning and model adaptation stages, the optimizer is Adam and the learning rate is $3\times 10^{-5}$ with the fixed numbers of epochs of $300$ and $50$, respectively. The probability of detection $P_{\mathrm{d}}$, the probability of false alarm $P_{\mathrm{f}}$, and the prediction accuracy $P_{\mathrm{acc}}$ are used as the performance metrics. The Parallel-CNN scheme in \cite{R. Mei}, the DeepSense scheme in \cite{D. Uvaydov} and the simultaneous orthogonal matching pursuit (SOMP) scheme in \cite{Y. Ma} act as baselines for performance comparison.

\begin{figure}[t!]
	\centering
	\includegraphics[width=3.2in]{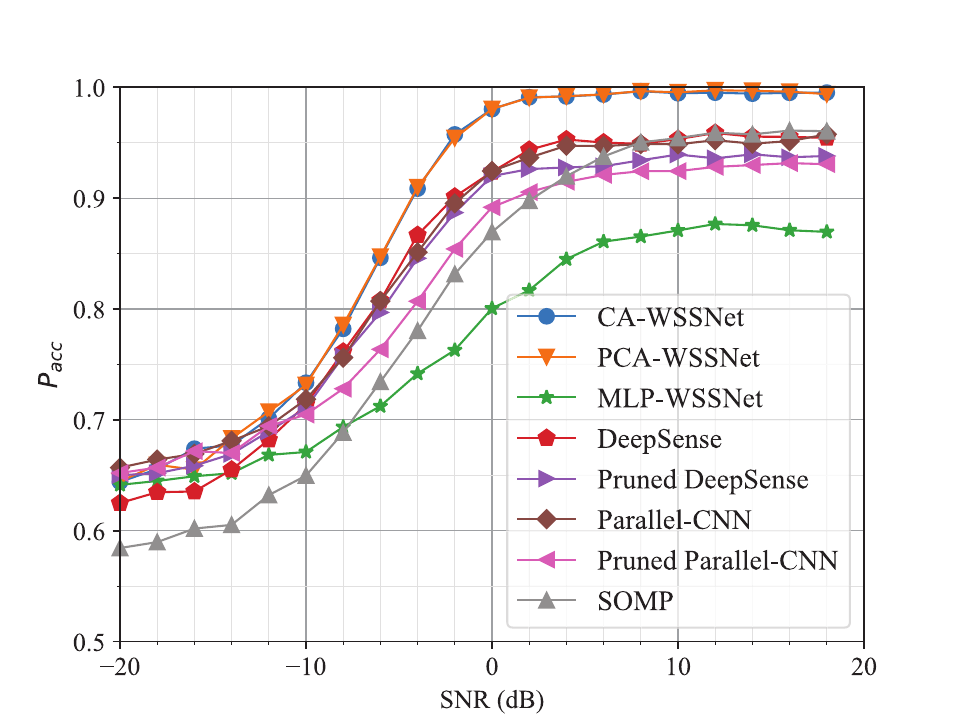}
	\caption{Prediction accuracy versus SNR for different WSS schemes with $K=12$.}\label{Pacc_vs_snr}
	\vspace{-0.2cm}
\end{figure}

\begin{table}[t!]
	\centering
	\caption{Comparison of Prediction Accuracy, Parameters, FLOPs, and Inference Time}
	\label{tab:method-compa}
	\resizebox{\linewidth}{!}{
		\begin{tabular}{ccccc}
			\toprule
			\makecell{Scheme} & \makecell{$P_{\mathrm{acc}}$} &\makecell{Parameters} & FLOPs & \makecell{Inference \\Time (ms)} \\
			\midrule
			PCA-WSSNet & \textbf{0.996}  & \textbf{136K} & \textbf{425K} & \textbf{0.089} \\	
			DeepSense &  0.958  & 660K & 13M & 0.1063 \\
			ParallelCNN & 0.949  & 165K & 1M & 0.1036 \\	
			\bottomrule
		\end{tabular}
	}
\vspace{-0.3cm}
\end{table}

Fig.~\ref{Pacc_vs_snr} shows the prediction accuracy versus signal-to-noise ratio (SNR) with the numbers of PUs $K=12$ for different schemes. Both the original and pruned versions of PCA-WSSNet, Parallel-CNN, and DeepSense are considered. For the original case, CA-WSSNet outperforms Parallel-CNN and DeepSense significantly while the latter two can achieve the performance no worse than SOMP. For the pruned case, PCA-WSSNet maintains the high performance stably while Parallel-CNN and DeepSense are degraded so obviously that SOMP surpasses them in the high SNR regime. In other words, the proposed PCA-WSSNet is robust to the weight pruning while Parallel-CNN and DeepSense should not be pruned to avoid the significant performance loss. Table~\ref{tab:method-compa} lists the comparison among PCA-WSSNet, Parallel-CNN, and DeepSense in terms of the prediction accuracy, number of parameters, FLOPs, and inference time with SNR $=8$ dB. Through the elaborated design of the network architecture and weight pruning, PCA-WSSNet is fully superior to Parallel-CNN and DeepSense. In addition, MLP-WSSNet excluding the CA module of PCA-WSSNet is also ploted as the ablation case. Compared with PCA-WSSNet, MLP-WSSNet loses about $12$\% accuracy at the high SNR region, demonstrating the effectiveness of the CA module.

\begin{figure}[t]
	\centering
	\includegraphics[width=3.2in]{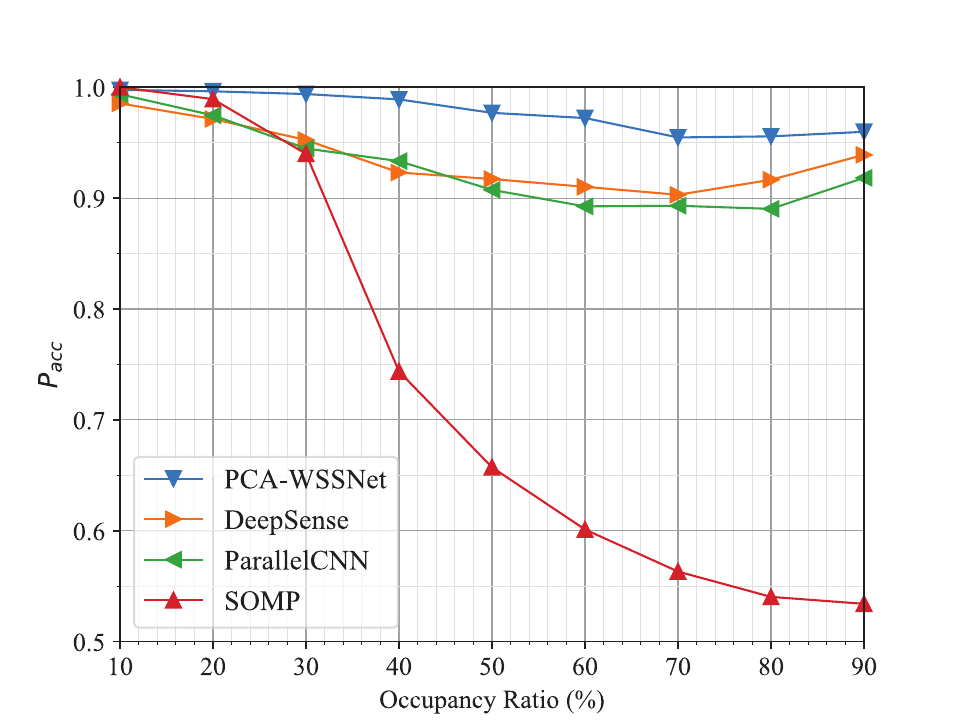}
	\caption{Prediction accuracy versus the occupancy ratio $r$ with SNR$=6$ dB.}\label{Acc_vs_occupancy}
\end{figure}

Fig.~\ref{Acc_vs_occupancy} shows the prediction accuracy versus the occupancy ratio $r$ with SNR$=6$ dB. From the figure, the performance of SOMP is high when $r \leq 20$\%, but becomes deteriorated as $r$ increases. In contrast, the other three DL-based approaches perform relatively steadily, where PCA-WSSNet achieves the best performance. For PCA-WSSNet with the fixed number of cosets $P=16$, its prediction accuracy is always higher than $0.95$ across all the considered values of $r$ even though the number of occupied sub-bands is larger than $P$, indicating that the proposed approach does not require the information of the number of occupied sub-bands as the prior knowledge.

\begin{figure}[t!]
	\centering
	\includegraphics[width=3.2in]{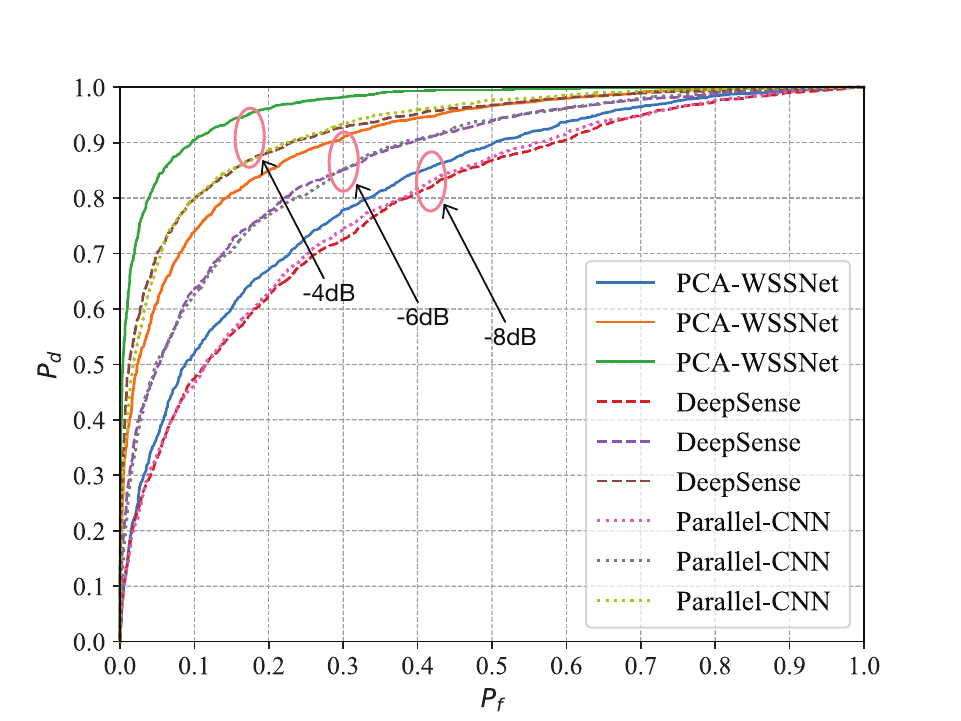}
	\caption{Probability of detection versus probability of false alarm with different SNRs.}\label{Pd_vs_Pf}
	\vspace{-0.3cm}
\end{figure}

Fig.~\ref{Pd_vs_Pf} shows the probability of detection versus the probability of false alarm with SNR=$-8$, $-6$, and $-4$ dB, which is exactly the receiver operating characteristic (ROC) curve to evaluate the DL-based solutions, i.e., PCA-WSSNet, Parallel-CNN, and DeepSense.\footnote{SOMP does not belong to the category of typical machine learning based methods and thus is excluded in this figure.} For each SNR, the ROC curve of PCA-WSSNet can completely encompass those of Parallel-CNN and DeepSense, thereby demonstrating the better generalization capability of PCA-WSSNet. Similar to Fig.~\ref{Pacc_vs_snr}, this performance superiority of PCA-WSSNet increases with SNR.

\begin{figure}[t!]
	\centering
	\includegraphics[width=3.2in]{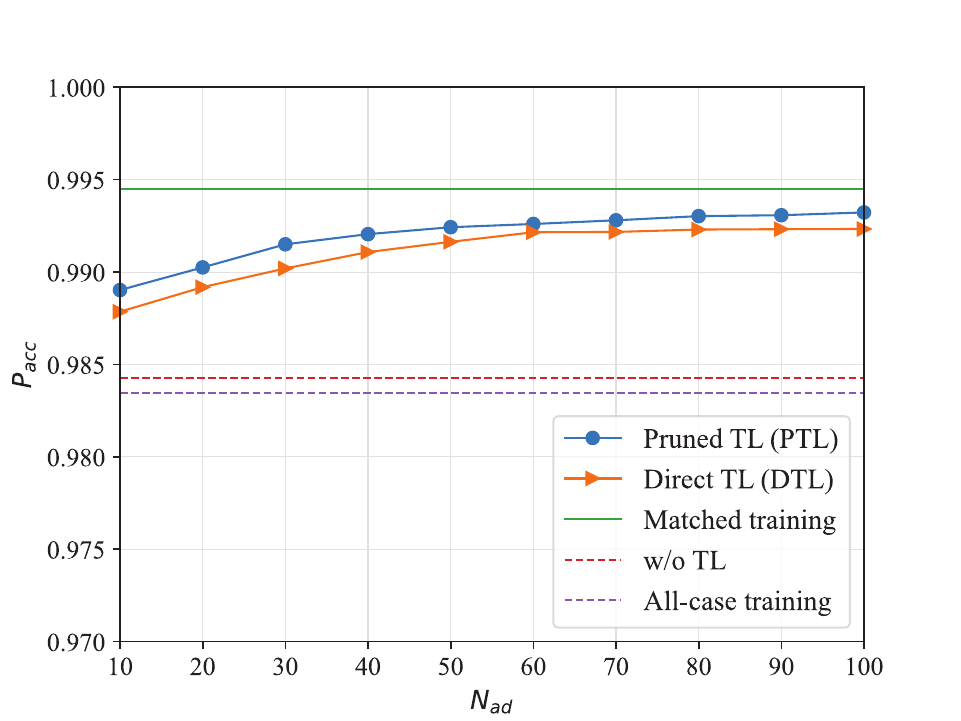}
	\caption{Prediction accuracy versus the number of adaptation samples with $K_{\mathrm{s}}=12$ and $K_{\mathrm{t}}=8$.}\label{Pacc_vs_Nad}
	\vspace{-0.3cm}
\end{figure}

To demonstrate how the selective weight pruning benefits the TL strategy, Fig.~\ref{Pacc_vs_Nad} show the prediction accuracy of the pruned TL (PTL) and direct TL (DTL) from $K_{\mathrm{s}}=12$ in the source domain to $K_{\mathrm{t}}=8$ in the target domain, where $N_{\mathrm{ad}}$ denotes the number of adaptation samples. Specifically, ``w/o TL" refers to using PCA-WSSNet trained with $K_{\mathrm{s}}=12$ to directly predict the spectrum situation with $K_{\mathrm{t}}=8$, ``Matched training" means training from scratch for the target domain $K_{\mathrm{t}}=8$, and ``All-case training" refers to including all considered occupancy cases in the offline training dataset. According to the ``w/o TL" curve, the prediction accuracy decreases significantly when the spectrum environment (the number of occupied sub-bands) changes, indicating the generalization problem plaguing PCA-WSSNet. ``All-case training" forces the DL model to compromise among many occupancy cases so that its performance is even worse than ``w/o TL". Compared with DTL, PTL improves the TL accuracy while reducing the number of weights, FLOPs, and inference time from $1$M, $2$M, and $0.105$ ms to $136$K, $425$K, and $0.089$ ms, respectively. Just needing $N_{\mathrm{ad}}=100$ adaptation samples fewer than $1$\% of the offline training set, PTL can retrieve most performance loss compared with ``w/o TL" and achieves the accuracy close to ``Matched training".

\vspace{-0.2cm}
\section{Conclusion}

In this article, PCA-WSSNet compatible with the sub-Nyquist sampling is proposed by organically integrating the MP and CA mechanism as well as embedding the selective weight pruning. TL strategy for PCA-WSSNet benefiting from the model pruning is developed to cope with the highly varied spectrum environment by adapting the domain-specific layers with a few adaptation samples. The proposed PCA-WSSNet outperforms the state of the art in terms of the sensing accuracy and complexity.



\ifCLASSOPTIONcaptionsoff
  \newpage
\fi

\end{document}